\documentclass[%
aip,
jcp,%
amsmath,amssymb,
reprint,%
]{revtex4-1}
\usepackage[english]{babel}
\usepackage[utf8]{inputenc}
\usepackage{graphicx}
\usepackage{dcolumn}
\usepackage{bm}

\newcommand{\ket}[1]{|#1\rangle}

\newcommand{\ob}[1]{{\langle #1\rangle}}

\begin{document}


\title[Non-Orthogonal Multi-Slater Determinant Expansions in Auxiliary Field Quantum Monte Carlo]{Non-Orthogonal Multi-Slater Determinant Expansions in Auxiliary Field Quantum Monte Carlo}

\author{Edgar Josu\' e Landinez Borda}
\affiliation{Lawrence Livermore National Laboratory, Livermore California 94550, United States}
\author{John Gomez}
\affiliation{Applied Physics Program and Department of Chemistry, Rice University, Houston Texas 77005-1892, United States }
\author{Miguel A. Morales}
\email{moralessilva2@llnl.gov}
\affiliation{Lawrence Livermore National Laboratory, Livermore California 94550, United States}

\date{\today}

\begin{abstract}
The Auxiliary-Field Quantum Monte Carlo (AFQMC) algorithm is a powerful quantum many-body method that can be used 
successfully as an alternative to standard quantum chemistry approaches to compute the ground state of
many body systems, such as molecules and solids, with high accuracy. In this article we use AFQMC with trial wave-functions
built from non-orthogonal multi Slater determinant expansions to study the energetics of molecular systems, including the 
55 molecules of the G1 test set and the isomerization path of the $[Cu_{2}O_{2}]^{2+}$ molecule.
The main goal of this study is to show the ability of non-orthogonal multi Slater determinant expansions to produce high-quality, 
compact trial wave-functions for quantum Monte Carlo methods. We obtain systematically improvable results as the number
of determinants is increased, with high accuracy typically obtained with tens of determinants. 
Great reduction in the average error and traditional statistical
indicators are observed in the total and absorption energies of the molecules in the G1 test set with as few as 10-20 determinants.
In the case of the relative energies along the isomerization path of the $[Cu_{2}O_{2}]^{2+}$, 
our results compare favorably with other advanced quantum many-body methods, including DMRG and complete-renormalized CCSD(T).
Discrepancies in previous studies for this molecular problem are identified and attributed to the differences in the number of electrons
and active spaces considered in such calculations. 
\end{abstract}

\maketitle

\section{Introduction}

Advances in the comprehension and predictive capabilities of electronic properties of matter, from single atoms to condensed matter systems, are a major quest that permeates many scientific and technological fields.
Due to the complexity of the fundamental equations of matter at the atomic scale, over the last several decades, computational methods have become a valuable tool in the discovery, characterization and optimization of new materials. First-principles computational methods, those that do not rely on empirical or experimental parameters and attempt a direct solution to the fundamental equations, have been mostly based on density functional theory (DFT)\cite{martin_2004}, due to its good predictive capability and modest computational cost. Unfortunately, DFT is based on approximations to electronic exchange and correlation which are known to be unreliable in many materials where these effects dominate or are difficult to approximate \cite{Cohen2012}, so called strongly correlated materials. As computer power increases and numerical algorithms improve, we are quickly approaching a point where the use of accurate quantum many-body approaches for the study of material properties is becoming feasible. Quantum many-body methods are typically orders of magnitude more computationally expensive than DFT, which has prevented their widespread application to bulk materials in the past, but could offer an accurate alternative with applicability even to strongly correlated materials. 

Traditional quantum chemistry methods, like Many-Body Perturbation Theory (MBPT), Coupled Cluster (CC) and Configuration Interaction (CI), can offer accurate solutions to the many-electron problem but their computational cost typically scales unfavorably with system size (N$^{6-7}$ for CC methods). While their extension to systems with periodic boundary conditions has been slow, implementations in standard computational packages are more common \cite{CRYSCOR,CP2K,VASP,PySCF} and applications to solids are appearing more frequently in the literature, including calculations based on second-order Moller-Plesser perturbation theory (MP2) \cite{marsmann2009,marsmann,vandevondele}, Random Phase Approximation \cite{paier2010,vandevondele2}, Coupled Cluster Singles-Doubles (CCSD) \cite{McClain2017}, among others. 

Quantum Monte Carlo (QMC) methods \cite{rmp2002} offer an important alternative to traditional quantum chemistry approaches for the study of many-electron problems, with both finite and periodic boundary conditions. They offer a favorable scaling with system size, typically between N$^3$-N$^4$, offer excellent parallel efficiency \cite{qmcpack1,qmcpack2}, and are capable of treating correlated electron systems with few approximations. Most QMC methods used in the study of realistic materials rely on a trial wave-function to control the notorious 
sign problem that plagues all fermionic Monte Carlo methods \cite{Troyer2005}. The trial wave-function not only controls the magnitude of the resulting approximation, but also the sampling efficiency and the magnitude of statistical uncertainties. Hence, accurate and efficient wave-function ansatz are important to the success of QMC methods in their application to realistic problems in physics, chemistry and material science.

In this article we examine non-orthogonal multi-Slater determinant (\textit{NOMSD}) expansions as an accurate and efficient trial wave-function ansatz for QMC simulations. We test the efficiency and accuracy of these wave-functions in combination with the Auxiliary-Field quantum Monte Carlo (AFQMC) method, as implemented in the QMCPACK \cite{qmcpack1,qmcpack2,qmcpack3,qmcpack4} simulation package. 
We show how these wave-functions have the capacity to systematically reduce errors associated with the phaseless approximation \cite{zhang3} in AFQMC, employed to control the sign problem, with highly compact expansions. These wave-functions have been used in the past to study strongly correlated lattice Hamiltonians with great success \cite{Fukome1,Fukome2,Tomita1,Imada1,Imada2}. They have also been recently popularized in connection with symmetry projection in Generalized Hartree-Fock theories \cite{scuseria1,scuseria2}. 
While this article focuses on calculations of small molecular systems, as a first application of the wave-function ansatz in AFQMC calculations of realistic Hamiltonians, similar improvements are expected when \textit{NOMSD} wave-functions are employed in other situations, including calculations of solids/extended systems, strongly correlated problems and other QMC approaches like Diffusion Monte Carlo. For example, we have successfully employed these wave-functions in studies of strongly correlated, periodic lattice Hamiltonians \cite{chia}. The \textit{NOMSD} wave-function ansatz now extends the arsenal of trial wave-functions employed in QMC calculations, including truncated CI expansions \cite{Morales,Clark}, Antisymmetric Geminal Powers (AGP)\cite{casula,neusscamman1,neusscamman2}, Pfaffians\cite{schmidth1,schmidth2}, Bardeen-Cooper-Schrieffer (BCS)\cite{gubernatis2,carlson}, among many others.

The structure of this paper is as follows: in the section II we briefly describe the wave function ansatz and its optimization method. Section III shows the application to the approach to the calculation of total and atomization energies for a subset of molecules of the G1 set. In section IV we describe the study of the isomerization path of the $[Cu_{2}O_{2}]^{2+}$ molecule, a scenario where different contributions to the correlation energy are significant along the path making the calculation quite challenging even for traditional methods. Finally, we discuss clear limitations with this approach and make concluding remarks.
 
\section{Non-Orthogonal Multi-Slater determinant Trial Wave Functions }

The trial wave-functions used in this work have the typical form:
\begin{equation}
  \ket{\Phi} = \sum_{i=1}^{n_d}\,c_i\,\ket{\varphi_i},
  \label{eq:FEDwf}
\end{equation}
where $\ket{\varphi_i}$ are Slater determinants and $c_i$ are linear variational parameters. In traditional quantum
Monte Carlo calculations, multi-determinants trial wave-functions of this form have been produced from truncated 
Configuration Interaction (CI) calculations \cite{Morales}, or more recently from selected CI calculations \cite{Caffarel12},
leading to expansions in orthogonal determinants connected by particle-hole excitations. While these expansions lead to systematically 
improvable results with reasonable stability and robustness, the expansions are typically very large requiring thousands of terms
in order to reach high accuracy \cite{Morales}. In this article, we propose the use of non-orthogonal Slater determinant expansions, 
where no orthogonality constraint is imposed between determinants, hence $\ob{\varphi_i|\varphi_j}\neq0$.
In fact, each Slater determinant is represented
as an orbital rotation from a given reference, $\ket{\varphi_i} = e^{\sum Z^i_{pq} c^\dagger_p c_q} \ket{\varphi_{ref}}$, where $Z$ 
is a Unitary matrix. 

The trial wave-function is generated using a version of the projected Hartree-Fock (PHF) algorithm developed
in the Scuseria group at Rice University \cite{Jimenez2012,scuseria1,scuseria3}.
Trial wave-functions are obtained by a direct minimization of the energy, $E=\ob{\Phi|\hat{H}|\Phi}/\ob{\Phi|\Phi}$, using a BFGS-like algorithm
and analytical energy gradients, see Jimenez-Hoyos, C., \emph{et al.}, \cite{scuseria3} for the relevant equations. 
We use 2 different approaches, 
the few-determinant (FED) algorithm\cite{scuseria1,scuseria2} and the resonating Hartree-Fock (ResHF) 
approach \cite{Fukome2,Fukome1,Tomita2}. In the FED algorithm, the Slater determinant expansion
is generated iteratively, adding and optimizing one determinant in each iteration to an already existing expansion.
During each iteration, determinants $\ket{\varphi_i}$  ($i=1,2,\ldots, n_d-1$) obtained from previous iterations are kept 
fixed\cite{scuseria1,scuseria2} and the energy is minimized with respect to the orbital rotation
matrix of the new determinant and all linear coefficients. This process continues until a given number of determinants is generated. At this point, the linear coefficients
are re-optimized by solving the associated eigenvalue problem. In the FED theory, symmetry projectors can be incorporated straightforwardly. The resulting 
single- or multi-reference symmetry-projected FED wave functions have been shown to be quite 
accurate. However, we will not focus on symmetry restoration in this work. 

In the ResHF approach, the energy is minimized with respect to all variational parameters in the trial wave-function, including
the rotation matrices of all determinants and all linear coefficients. In this work, we use a combination of both approaches to generate our trial wave-functions. 
We first generate a NOMSD expansion of a given length (e.g. 20 determinants) using the FED approach. The resulting expansion is then optimized 
using the ResHF approach, but without new determinants. In other words, we use the FED approach to generate an initial set of parameters for the ResHF
approach. We find this to be successful in most situations studied so far. In some situations, particularly when generating long expansions, we insert ResHF steps
in between blocks of FED steps to improve the stability and efficiency of the optimization at later stages of the iterative cycle. We must mention that we almost never
converge the ResHF calculations to high precision. This is typically expensive to do, but also not necessary in AFQMC. The trial wave-function in AFQMC does
not need to satisfy any stationary properties, so there is no need for high convergence.

The mean-field orbitals used in the FED and ResHF theories could be obtained from restricted-HF (RHF), unrestricted-HF (UHF), 
or even generalized-HF (GHF) wave functions. The three cases represent different levels of symmetry-breaking of the Hartree-Fock determinant. 
In this work, we use orbitals from RHF wave-functions in closed-shell systems and from UHF in open-shell cases.
While the use of GHF wave-functions can lead to significant improvements in our calculations in combination with NOMSD, 
we have not explored this in this work and will be the focus of a future publication.

\begin{figure}[h!]
\includegraphics[width=0.48\textwidth]{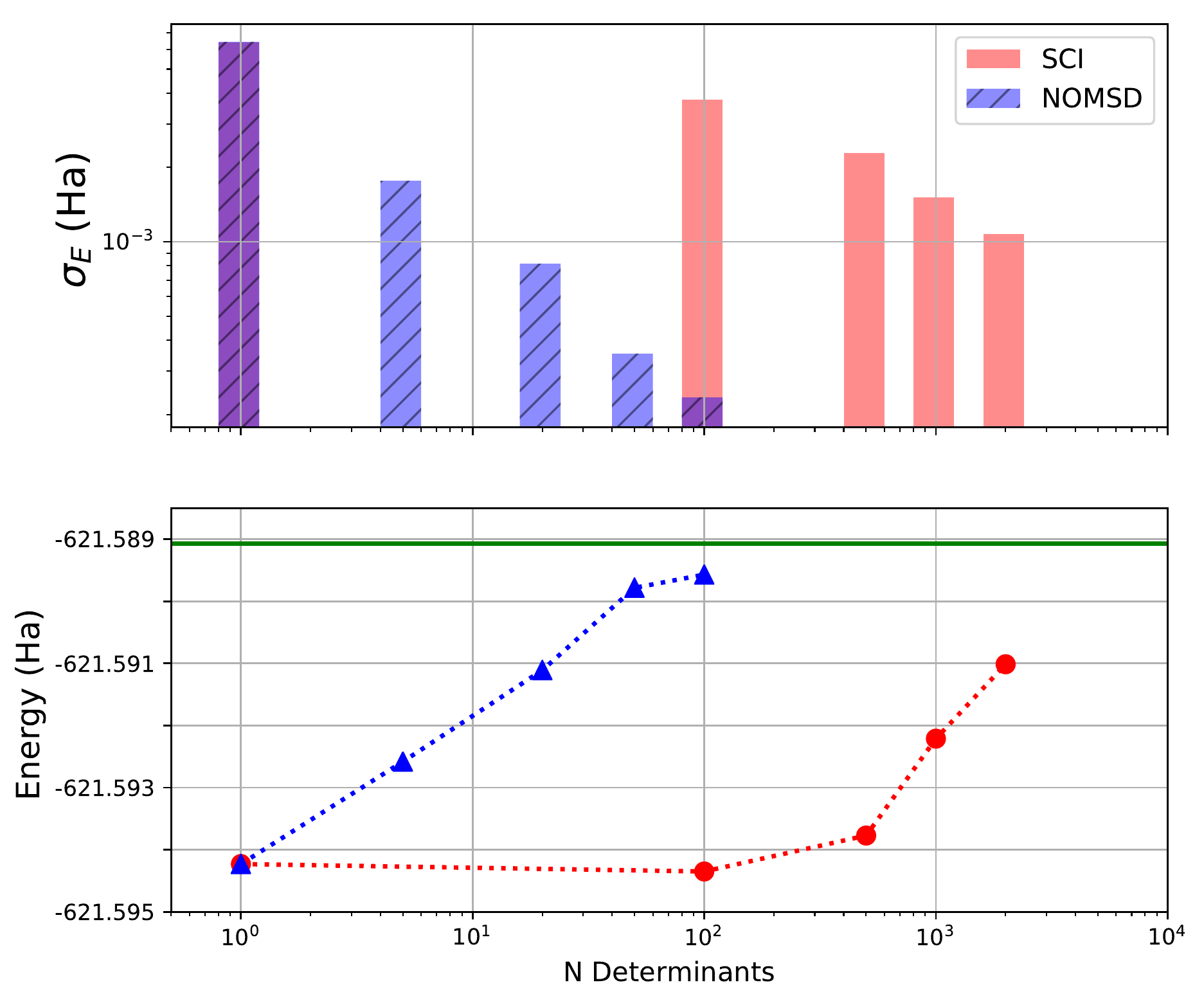}
\caption{Performance comparison of multi-determinant trial wave functions in AFQMC calculations for the $NaCl$ molecule 
at the equilibrium geometry with a cc-pVDZ basis set. 
Results for a \textit{NOMSD} expansion are shown in blue (dashed bars on top and triangles in bottom figures) and for a
orthogonal expansion obtained from a selected Configuration Interaction calculation shown red (clear bars on top and circles in bottom figures).
The top figure shows the standard deviation of the energy estimator, while the bottom figure shows the total energy, both as a function of the number of
determinants in the expansion.The horizontal line in the bottom figure indicates the CCSDTQ result.
Orthogonal expansions require expansions with approximately 2 orders of magnitude more terms to reach similar values of the energy and standard deviation.
Error bars are smaller than the symbol size.}
\label{fig:wfnconv}
\end{figure}

Figure \ref{fig:wfnconv} shows a comparison of the performance of AFQMC calculations for the $NaCl$ molecule using a \textit{NOMSD} wave-function and an orthogonal multi-Slater determinant
expansion obtained from selected Configuration Interaction (selCI), for which we used the DICE code \cite{Holmes16,Sharma17}. The \textit{NOMSD} wave-function requires approximately 2 orders of
magnitude less determinants in the expansion to reach a similar reduction in phase-less bias and energy fluctuations compared to the selCI wave-function. We have found this to be the case
in all situations we have studied so far. 
Since actual execution times will be very sensitive to implementation details and code optimization (for example, low rank update schemes \cite{Clark} can be used to speed up the evaluation of orthogonal expansions), we have decided to present such comparisons and details of the implementation in a future publication focused on those aspects of the calculation.
From the results presented above, it is clear that the \textit{NOMSD} approach leads to a compact and systematically improvable wave-function which can succesfully compete with state-of-the-art alternatives in the field.

\section{Simulation details}

The MOLPRO\cite{molpro} software package was used to perform all Hartree-Fock and CCSD(T) calculations. In addition, it was also used to generate all the integral files (matrix elements of the Hamiltonian in the HF basis) necessary for both AFQMC calculations and for the iterative Hartree-Fock method used to generate the \textit{NOMSD} wave-functions. The PHF code\cite{Jimenez2012,scuseria1,scuseria3} was used to generate \textit{NOMSD} wave-functions. CCSDTQ calculations were performed with the AQUARIUS package \cite{aquarius}. All AFQMC calculations were performed with the QMCPACK software package \cite{qmcpack4}. QMCPACK offers a general implementation of the AFQMC algorithm, including standard improvements to the method like mean-field subtraction \cite{PhysRevE.70.056702}, force bias \cite{PhysRevE.70.056702}, hybrid propagation \cite{Purwanto09}, single and multi-determinant wave-functions with both orthogonal and non-orthogonal multi-Slater determinants, population control, large scale parallelization, multi-core implementations, among many others. The code can be used to study both finite as well as extended systems. Interfaces to the code exists for Molpro, GAMESS, PySCF and VASP, the latter 2 codes able to generate input for calculations with periodic boundary conditions. Both free projection \cite{Motta16} and propagation with the phase-less constraint \cite{Zhang2003} are implemented. For additional implementation details we refer the reader to the QMCPACK reference article \cite{qmcpack4} and to the code's manual \cite{qmcpack3}. For additional details about the AFQMC method, we refer the reader to the many excellent articles in the literature \cite{Zhang2003,PhysRevE.70.056702,Al-Saidi06,Purwanto09,Motta16}.

All calculations presented in this article employed the phase-less approximation \cite{Zhang2003}. Single determinant trial wave-functions were generated with Restricted Hartree-Fock for closed-shell systems and with Restricted Open-Shell Harthree-Fock for open shell cases. Unless otherwise stated, 
we employed a time-step of 0.005 $ha^{-1}$ and 576 walkers, both choices leading to systematic errors below $\simeq$ 0.2 mHa. Due to the complexity of some of the figures, 
we typically avoid plotting error bars to improve the clarity of the data presented. Unless otherwise stated, all error bars have been converged to less than 0.4 mHa in single determinant
calculations (average error of  $\simeq$ 0.2 mHa) and to less than  $\simeq$ 0.2 mHa in \textit{NOMSD} calculations (average error below  $\simeq$ 0.1 mHa).
All calculations employ the frozen-core approximation. We chose the standard cores used by MOLPRO, namely He cores are frozen in 1st row atoms while Ne cores are frozen in the case of 2nd row atoms. For a detailed analysis of the influence of frozen-core approximations on the total and atomization energies, see Ref. \cite{Feller4}. 
Correlation-consistent atomic basis sets were used and geometries were taken from Ref. \cite{Grossman} for the molecules in the G1 test set and from Ref. \cite{scuseria} for the study of the isomerization path of the copper oxide molecule.

\section{G1 Test Set}

In this section we study the energies of 55 molecules of the G1 test set \cite{Curtiss} to analyze the accuracy and rate of convergence of the AFQMC method when \textit{NOMSD} trial wave-functions are used. This set has a long history in both quantum chemistry and quantum Monte Carlo communities, often 

used as benchmark to test new developments and capabilities \cite{Grossman,Towler,Morales}. Accurate reference data exists for all molecules in the set, including atomization energies \cite{Feller4}. A large literature exists with careful analysis of the influence and magnitude of the various approximations employed in theoretical calculations \cite{Feller4,Feller3,Feller2,Feller1}, including effects associated with the choice of basis set and approximations to electronic correlation. 

\subsection{Total Energies}

\begin{figure}[h!]
\includegraphics[width=0.45\textwidth]{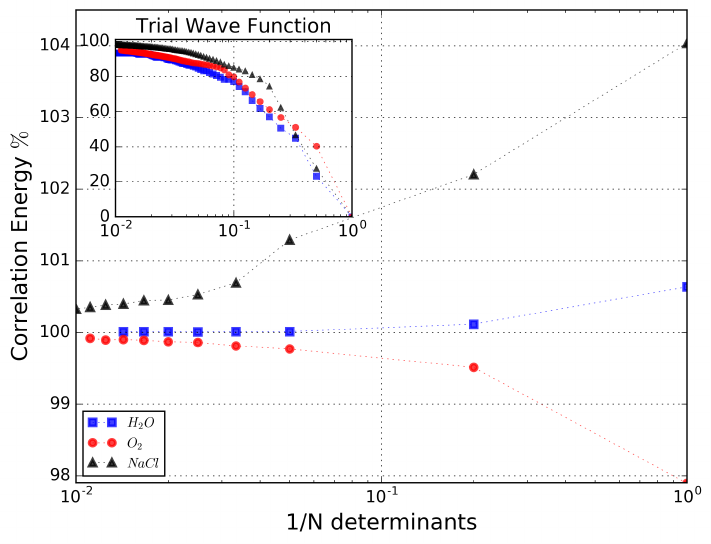}
\caption{Convergence of AFQMC correlation energy with the number of determinants in a non-orthogonal expansion.
Equilibrium geometries and a cc-pVDZ basis set has been used in all cases.
The inset shows the convergence of the correlation energy of the trial wave function.}
\label{fig:phfcov}
\end{figure}

As an illustrative example, Figure \ref{fig:phfcov} shows the percentage of the correlation energy (measured with respect to CCSDTQ), obtained by AFQMC, as a function of the number of determinants in the \textit{NOMSD} expansion, for the molecules $H_2O$, $O_2$, and NaCl. This figure shows several features of the use of \textit{NOMSD} wave-functions in AFQMC. First notice that, since all 3 molecules possess weakly correlated electronic structures, single determinant calculations already produce reasonable results with errors around $\sim$1$\%$-4$\%$. In all cases the amount of correlation energy approaches systematically $100\%$ as the determinant expansion in increased, even though the convergence can be from above since AFQMC is not a variational method. In the case of $H_2O$ and $O_2$, less than 10 determinants are needed to obtain ~99.5$\%$ of the correlation energy, while at least 50 determinants are needed to obtain a comparable accuracy in NaCl. As will be seen below, molecules with a strong ionic character tend to have a slower convergence rate with respect to the number of determinants. Nonetheless, even though the specific rate of convergence clearly depends on the specific molecule, rapid convergence is observed in all three cases. 

\onecolumngrid

\begin{figure}[htb!]
\includegraphics[scale=0.165]{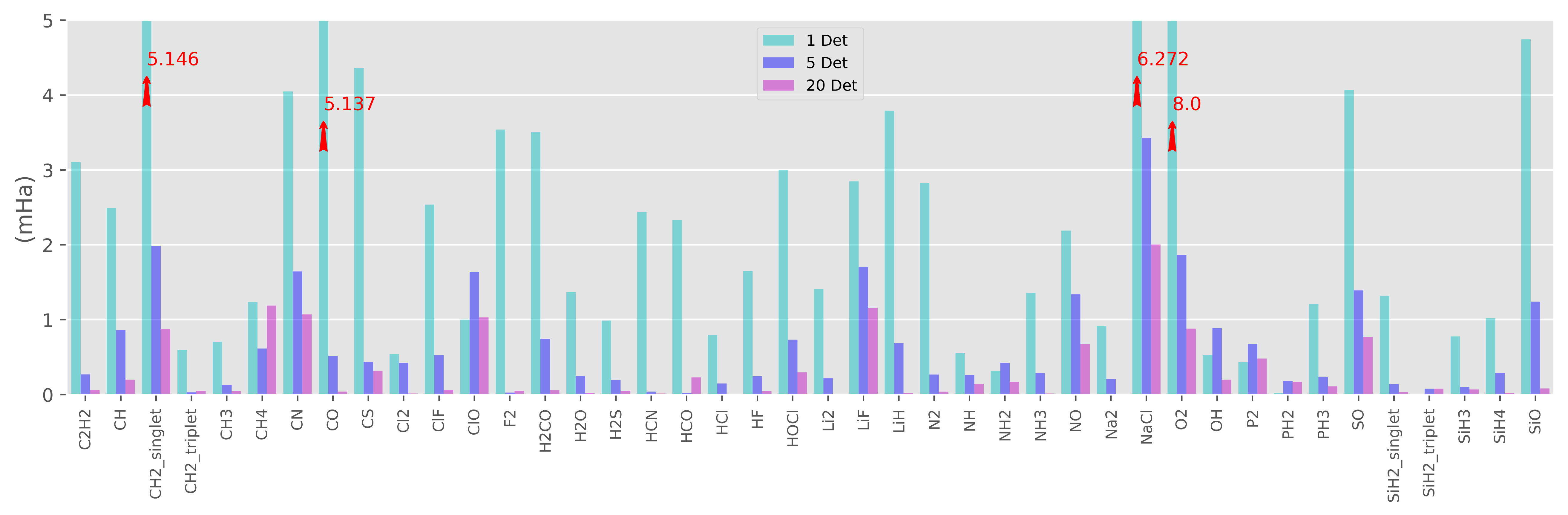}
\caption{Comparison of AFQMC total energies differences, employing \textit{NOMSD} trial  wave functions with 1,5 and 20 determinants, against Coupled-Cluster calculations. The Coupled-Cluster calculations include single, double, triple and quadruple excitations (CCSDTQ). A cc-pVDZ basis set has been used in all cases.
The red arrows indicate that the values of that difference are greater than 5 mHa and point to its explicit value}
\label{fig:totalE}

\end{figure}

\onecolumngrid

\begin{table}[htb!]
\begin{ruledtabular}
\caption {Raw performance statistics of the difference of the total energies between AFQMC with the NOMSD expansion and
CCSDTQ. All calculations were performed with Dunning's cc-pVDZ basis set and employed the frozen-core approximation. Column definitions -Max. neg.: maximum negative difference; Max. pos: Maximum positive difference; ME: mean error; MAE: mean absolute error; STDEV: standard deviation. All quantities are in mHa.}
\begin{tabular}{cddddd}
AFQMC $N_{det}$ & \mbox{Max. neg.} & \mbox{Max. pos.} & \mbox{ME} & \mbox{MAE} & \mbox{STDEV}\\
\hline
1 det & -8.00 & 6.27 & 1.01 & 2.27 & 2.72 \\ 
5 det & -1.8 & 3.42 & 0.11 & 0.65 & 0.95\\
20 det & -1.07 & 2.00 & -0.03 & 0.31 & 0.54
\end{tabular}
\end{ruledtabular}
\label{tab:total}
\end{table}

\twocolumngrid

In order to have a more quantitative characterization of the accuracy of the AFQMC-\textit{NOMSD} approach, we performed CCSDTQ calculations for a subset of the 55 molecules considered in this work using Dunning's \cite{Dunning} cc-pVDZ basis set. We were forced to limit our analysis to a subset of the 42 molecules in the test set due to difficulties with the convergence of CCSDTQ for some of the molecules. 
We expect CCSDTQ to provide reliable results for these molecules, with errors below 1 mHa \cite{Feller4}. Figure \ref{fig:totalE} shows the error in the total energy calculated by AFQMC-\textit{NOMSD} when measured with respect to CCSDTQ. We present results using 1, 5, and 20 determinants in the \textit{NOMSD} expansions to show the convergence of the correlation energy as the expansion is increased. Table \ref{tab:total} shows the associated performance statistics for the energy differences. For the selected set of molecules, the mean absolute error decreases from 2.27 mHa to 0.31 mHa by going from single determinant to a 5 determinant expansion, with a further reduction to 0.34 mHa when the expansion is increased to 20 determinants. 
In general, all statistical measures of the energy difference show a significant and systematic reduction between single determinant calculations and NOMSD with increasingly larger expansion sizes. 
Notice that the choice of a 20 determinant expansion is somewhat arbitrary, with the intention of showing a very significant reduction on the errors with just a handful of determinants. 
While the magnitude of the errors is dependent on the molecule, the capacity for significant improvement with a very compact expansion is universally observed. 

\subsection{Atomization Energies}

\begin{figure}[htb!]
\includegraphics[scale=0.165]{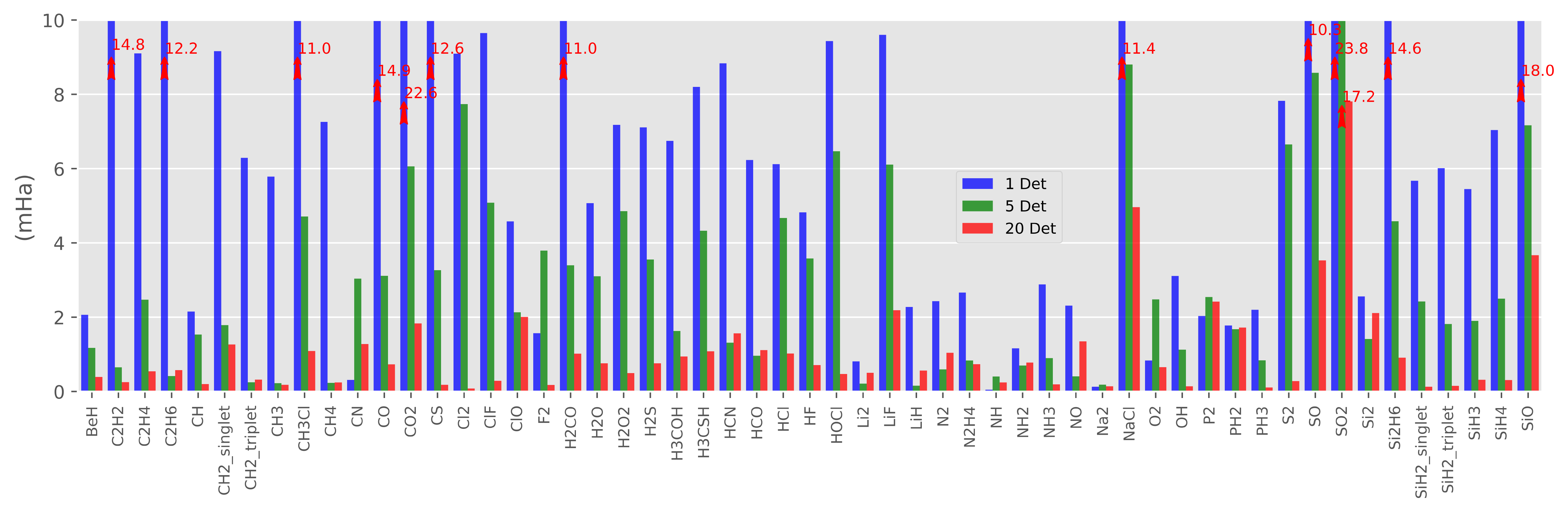}
\caption{Atomization energies for the set of molecules, comparison of the extrapolated data of AFQMC for the cc-pVXZ, $X=\{D,T,Q\}$, displaying the mean absolute error
in comparison to the reference values}
\label{fig:atomization}
\end{figure}

Figure \ref{fig:atomization} shows the error in the atomization energy, calculated with AFQMC-\textit{NOMSD}, for the 55 molecules considered in this work. Table II 
shows the associated performance statistics. Reference values have been obtained from Feller \emph{et al.} \cite{Feller3,Feller4} by subtracting the Zero Point Energy (ZPE), Spin Orbit (SO), Core Valence (CV), Scalar Relativistic (SR) and the Diagonal Born-Oppenheinmer Correction (DBOC) contributions from the experimental values.  Atomization energies have been extrapolated to the Complete Basis Set (CBS) limit using results with Dunning's cc-pV(X)Z basis sets, with $ X=\{D,T,Q\}$. We used the mixed Gaussian/Exponential formula, $E(n) = E_{CBS} + A e^{-(n-1)} + B e^{-(n-1)^2}$, which has been shown to perform well for CCSD(T) \cite{Feller1, Feller2}. 
Similar to the previous section, we show results for expansions of size 1, 5, and 20, with the goal of showing the capacity of this approach for error reduction with short, compact expansions. 

\onecolumngrid

\onecolumngrid
\begin{table}[htb!]
\begin{ruledtabular}
\caption {Raw performance statistics of the AFQMC atomization energies error with the \textit{NOMSD} expansion respect to the
reference values. All quantities are in mHa.}
\begin{tabular}{cddddd}
AFQMC $N_{det}$ & \mbox{Max. neg.} & \mbox{Max. pos.} & \mbox{ME} & \mbox{MAE} & \mbox{STDEV}\\
\hline
1 det & -23.78 & 2.66 & -6.32 & 6.68 & 5.78\\
5 det & -17.15 & 3.04 & -2.93 & 3.38 & 3.59\\
20 det & -7.81 & 2.42 & -0.52 & 1.17 & 1.78
\label{tab:atom}
\end{tabular}
\end{ruledtabular}
\end{table}

\twocolumngrid
As expected from the discussion above, the \textit{NOMSD} expansion is able to reduce significantly the errors of AFQMC throughout the test set. 
Overall, the mean absolute error is reduced from 6.68 mHa to 3.38 mHa when going from a single determinant to a 5 determinant expansion, and to 1.17 mHa when 20 determinants are used. 
Several observations can be made upon careful inspection of the various molecules in the set. 
First, we notice that for many molecules a 5 determinant expansion is enough to drastically reduce the error in the atomization energies, 
often with error reductions of over 90$\%$. We can also notice a set of molecules for which convergence is slow, 
including for example LiF and NaCl. 
These molecules have a strong ionic character, which seems to lead to slow convergence with respect to expansion length. 
We must point out that the errors in the atomization energies discussed above contain, 
in addition to any bias associated with the AFQMC method, 
a contribution from basis set extrapolation since the largest basis set we considered was cc-pVQZ. 
Since the focus of the article is the characterization of the \textit{NOMSD} ansatz, we decided to avoid
a careful analysis of basis set extrapolation errors in the results presented below. For a detail analysis of basis set 
effects in atomization energies of the G1 test set, we refer the reader to Ref. \cite{Feller2}. 

\section{Copper Oxide Isomerization Path}

\begin{figure}[!hbt]
\flushleft
\includegraphics[scale=1.3 ]{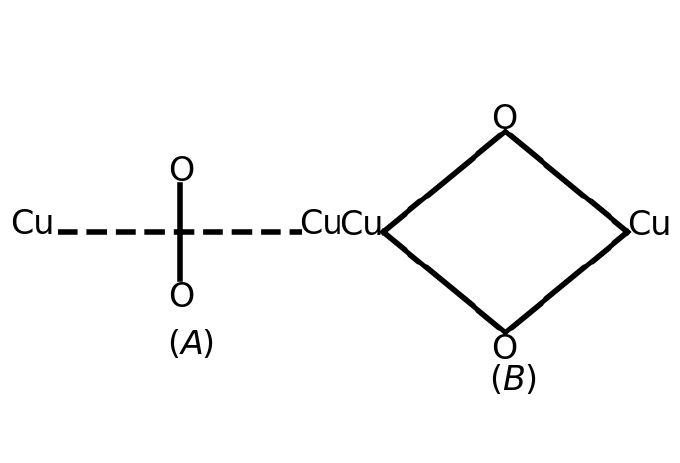}
\caption{Schematic representation of the geometry of the $[Cu_{2}O_{2}]^{2+}$ molecule at the endpoints the isomerization path. Configuration \textbf{A} ($f=100$) is characterized by atomic distances $R_{OO}\simeq 1.4\AA$ and $R_{CuCu}\simeq 3.6\AA$, while configuration \textbf{B} ($f=0$) is characterized by distances $R_{OO}\simeq 2.3\AA$ and $R_{CuCu}\simeq 2.8\AA$.
Similar to previous studies, a linear interpolation between the two configurations defines the path.
} 

\label{fig:iso}
\end{figure}

The study of the isomerization path of the $[Cu_{2}O_{2}]^{2+}$ molecule has become a common practice in the last decade
to test the capacity of ab initio and quantum chemistry methods to accurately describe the energetics of correlated molecular problems. As pointed out by K. Samanta \emph{et al.}, \cite{scuseria}, the interest comes from the dissimilar nature of the correlation energy in the molecule through the path. A successful description of the relative energy of the molecule along the path requires methods that offer a balanced treatment of both static and dynamic components of the correlation energy, since both terms have dominant contributions at different points along the path.

Figure \ref{fig:iso} shows a schematic representation of the stable isomers at the end points of the isomerization path. Following previous studies, we define a linearized isomerization path between isomers A and B through the quantity \textit{f} defined by the equation $q_{i} (f)= q_{i}(f=0)+\frac{f}{100}(q_{i}(f=100)-q_{i}(f=0))$; where $q_{i}(f)$ are the coordinates of the atoms in the molecule along the isomerization path, $q_{i}(f=0)$ are the atomic coordinates for isomer B and $q_{i}(f=100)$ are the atomic coordinates for isomer A. 
We use the same geometry, pseudopotential and basis set employed in several previous studies of this molecule in order to be able to draw a direct comparison with other correlated electronic structure approaches. For a detailed description and a summary from previous theoretical calculations, we refer the reader to Refs. \cite{Gherman09}, \cite{scuseria} and references therein.

Previous studies of this molecule using advanced electronic structure methods including Density Matrix Renormalization Group (DMRG) \cite{Yanai}, Complete Renormalized Coupled Cluster (CR-CC)\cite{Cramer}, and Projected Hartree-Fock (PHF)\cite{scuseria}, appear to lead to discrepancies on the relative energy differences across the isomerization path. In particular, PHF calculations reported in Ref. \cite{scuseria} seem to lead to larger energy differences when compared to DMRG and CR-CC. As shown below, these discrepancies seem to be related to the different choices of core and active spaces selected in each of the calculations. While all these calculations appear to use the same atomic basis set and the same molecular geometries, DMRG calculations reported in Ref. \cite{Yanai} used an active space of (28e,32o) leading a calculation with 28 correlated electrons, CR-CC calculations correlated explicitly the 2s and 2p orbitals of oxygen in addition to the 4s and 3d orbitals of Cu leading to 32 correlated electrons, while PHF calculations in Ref. \cite{scuseria} correlated all 52 electrons in the calculation. As will be shown, the resulting differences in the number of core electrons in each of the calculations leads to the observed discrepancies in previous results, rather than strong inaccuracies in the predicted relative energies. 

\begin{figure}[!hbt]
\includegraphics[width=0.5\textwidth, angle=0 ]{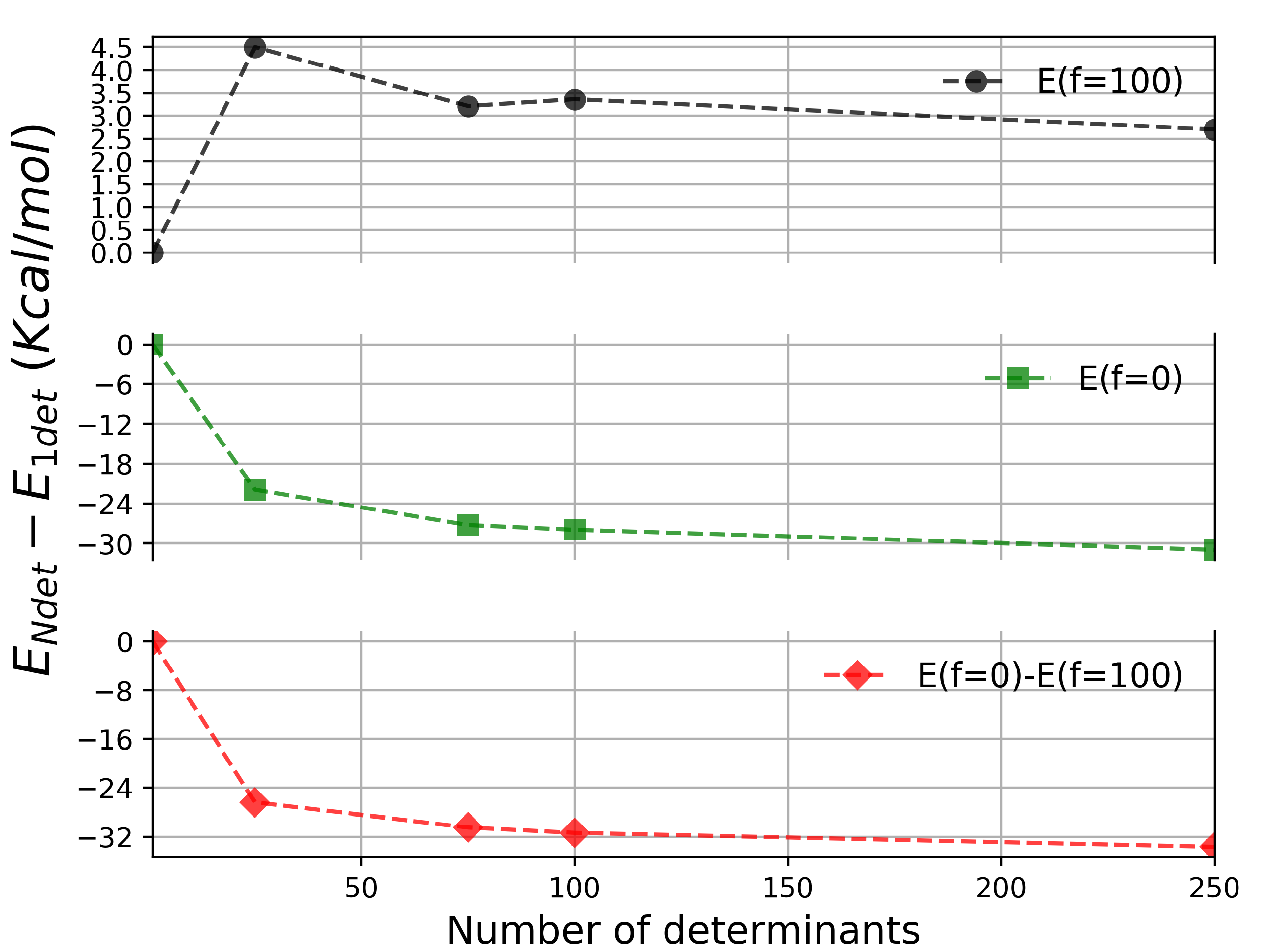}
\caption{Convergence of the AFQMC-\textit{NOMSD} total energy for the $[Cu_{2}O_{2}]^{2+}$ molecule as a function of the number of determinants in the expansion for the endpoints of the isomerization path. }
\label{fig:cu2o2_afqmc}
\end{figure}

Figure \ref{fig:cu2o2_afqmc} shows the dependence of the AFQMC-\textit{NOMSD} total energy with the number of terms in the expansion for the 2 endpoints of the isomerization path of the $[Cu_{2}O_{2}]^{2+}$ molecule. 

In this case 32 electrons (CR-CC) are directly correlated in the calculation. 
As we can see, single determinant trial wave-functions lead to poor results for this molecule with errors in the relative energy on the order of 35 (Kcal/mol). 
Convergence in the relative energy between the endpoints of the path are reached with \textit{NOMSD} expansions with several hundred determinants. 
This is to be expected since the electronic structure of configuration \textbf{B} has significant static correlation, 
requiring expansions with over 200 determinants for reasonable convergence. 
The total energy of configuration \text{A} converges quickly, requiring only 75 determinants for an accuracy of 1 kcal/mol, 
since in this case dynamic correlation is dominant and is efficiently captured by the AFQMC method. 
Even in this complicated scenario, \textit{NOMSD} offers a systematically convergent, compact choice for trial wave-functions.

\begin{figure}[!hbt]
\includegraphics[width=0.5\textwidth, angle=0 ]{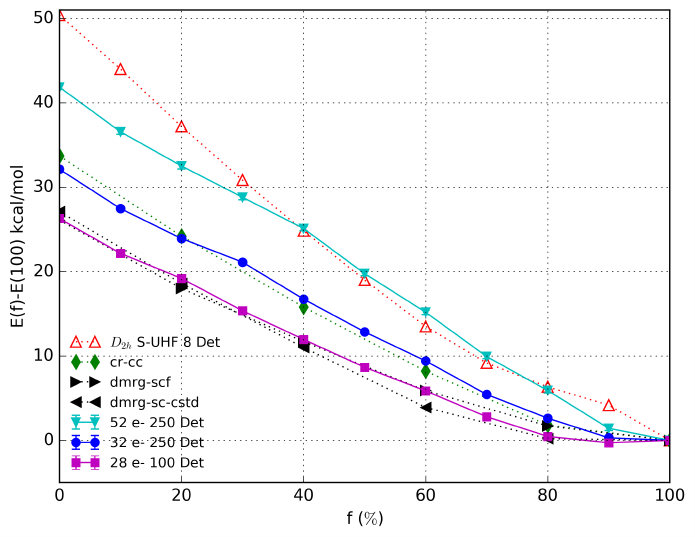}
\caption{Relative energy of the isomerization path comparing \textit{NOMSD}-AFQMC with previous results.}
\label{fig:cu2o2}
\end{figure}

Figure \ref{fig:cu2o2} shows the total energy of the molecule along the path, relative to configuration \textbf{B}. 
We show results for AFQMC-\textit{NOMSD} calculations with 28, 32 and 52 correlated electrons in order to mimic the choices made by previous calculations, which we also show. 
AFQMC calculations with 28 active electrons used 100 determinants, while calculations with 32 and 52 active electrons used 250 determinants. 
Larger active spaces required larger expansions in order to reached the similar levels of accuracy. 
Excellent agreement is observed between AFQMC, CR-CC and DMRG, when the appropriate number of active electrons is considered.  
This is a very strong indication that the discrepancies between these methods is mostly attributable to the differences in the number of correlated electrons. 

\section{Discussion}

Several observations about the \textit{NOMSD} approach are appropriate at this point. First we must point out that, similar to truncated CI expansions, the \textit{NOMSD} expansion is not size extensive nor size consistent. 
This means that the number of terms in the expansion necessary for an equivalent description of the system will ultimately scale exponentially as a function of the system size. 
While this creates serious problems for direct use of truncated CI (or non-orthogonal)  expansions in quantum chemistry methods for periodic systems, 
the relevant question in our case is not whether the wave-function is size-extensive but whether the reduction of the phase-less error is. While a study of the scaling of the phase-less error with expansion length is outside the scope of the work presented in this article, it is unlikely that in the thermodynamic limit non-exponential behavior is obtained. Nonetheless, for systems of typical interest (up to ~100 atoms), this approach leads to a simple and well-defined path for error reduction. For larger calculations, it is possible to employ the \textit{NOMSD} approach in an active space approach, where only a small subset of the orbitals are included in the construction of the correlated trial wave-function. This will lead to better scaling and higher efficiency without sacrificing too much accuracy. A future publication will discuss modifications and improvements of the method needed in its application to correlated materials in periodic boundary conditions. 

Another important consideration is the evaluation efficiency of the \textit{NOMSD} wave-function compared to traditional expansions based on truncated or selected CI. Since different configurations in orthogonal expansions differ by a finite number of orbitals, fast low-rank update schemes can be used \cite{Clark} which leads to highly efficient evaluation routines. In the case of \textit{NOMSD} expansions, no such optimizations have been devised leading to much higher evaluation cost as a function of expansion length. A proper comparison of the two methods requires a careful analysis of implementation details, which is outside the scope of this work. Nonetheless, we believe that the current approach offers sufficiently compact expansions to compensate for the lower evaluation efficiency. A detailed analysis of the computational aspects of the method will be presented in a separate publication.

\section{Conclusions}

The AFQMC method offers a powerful alternative for the study of quantum many-body problems in physics, chemistry and material science. Its ability to handle both weakly and strongly correlated electronic structure problems with favorable scaling offers an excellent alternative to traditional methods in quantum chemistry and ab initio electronic structure. When combined with flexible and efficient trial wave-functions, the method has the potential to become the method of choice for the study of realistic quantum systems. In this article, we present non-orthogonal Multi-Slater determinant expansions as an accurate and compact choice for trial wave-functions in AFQMC. They offer the flexibility, simplicity and compact representation needed for realistic calculations of materials and complicated molecular systems. \textit{NOMSD} expansions have the capability to recover significant fractions of the correlation energy with a modest number of terms, compared to traditional orthogonal expansions based on configuration interaction which typically have slowly decaying tails which require orders of magnitude more terms for similar accuracy. 

For weakly correlated systems, including the molecules in the G1 test set, 
we showed the capacity of the \textit{NOMSD} approach to systematically reduce the errors in total and atomization energies with short expansions, 
even as small as 5 determinants in many cases. We obtain a mean absolute error of 1.17 mHa in the atomization energies of the 55 molecules considered 
in this work with a 20 determinant expansion, which includes potential errors associated with basis set extrapolation. 
In addition, we obtain an mean absolute error of 0.31 mHa in the total energies when comparing against CCSDTQ, 
for a subset of the 55 molecules in the cc-pVDZ basis. These errors can easily be reduced further with longer expansions if desired.

For the correlated molecular system, $[Cu_{2}O_{2}]^{2+}$, 
we showed the vast improvement obtained in the relative energies along the isomerization path from the use of \textit{NOMSD} expansions in combination with AFQMC. 
In this case, AFQMC calculations with single determinant trial wave-functions lead to large errors in the relative energies along the path, 
with errors as large as 30 kcal/mol when compared to advanced electronic structure methods like DMRG and CR-CC. 
In this case, \textit{NOMSD} expansions of several hundred determinants were necessary in the AFQMC calculations to obtain well converged results. 
We also showed that discrepancies in previous calculations can be well explained by the differences in the number of core electrons used, 
obtaining excellent agreement with CR-CC and DMRG when consistent choices are made. 
We believe that the work presented in this article will lead to a further examination and optimization of the use of non-orthogonal 
compact expansions in quantum Monte Carlo and will further contribute to the quest for accurate and efficient wave-function ansatz for correlated quantum many-body systems.

\section{Acknowledgements}
We would like to thank Chia-Chen Chang, Matthew Otten and Gustavo E. Scuseria for useful discussions.
We would also like to thank Carlos Jimenez-Hoyos for providing us access to the PHF code used to generate \textit{NOMSD} 
expansions.
The PHF code used in this
work was developed in the Scuseria group at Rice University \cite{Jimenez2012,scuseria1,scuseria3}.
The contribution of John Gomez to this work was carried out during
his summer internship at LLNL in 2016. 
This work was performed under the auspices of the U.S. Department of Energy by Lawrence Livermore
National Laboratory and supported by LLNL-LDRD project No.\ 15-ERD-013.
Livermore National Laboratory is operated by Lawrence Livermore National
Security, LLC, for the U.S. Department of Energy, National Nuclear Security Administration
under Contract DE-AC52-07NA27344. 
J. A. G. acknowledges the support of the National Science Foundation Graduate Research Fellowship Program (DGE-1450681).

\bibliographystyle{unsrtnat}
\bibliography{datav3}

\end{document}